\documentclass[
  aps,
  prl,
  reprint,
  amsmath,amssymb,
  floatfix,
  superscriptaddress
]{revtex4-2}
\usepackage{xcolor}
\usepackage{graphicx}
\usepackage{orcidlink}
\usepackage{bm}
\usepackage{hyperref}
\hypersetup{colorlinks=false,pdfborder={0 0 0}}

\begin{document}

\title{Programmable Hybrid Exceptional Points in Passive Scattering Networks}

\author{Kaiyuan Wang\orcidlink{0009-0001-1171-3052}}
\affiliation{School of Electrical and Electronic Engineering, Nanyang Technological University, 50 Nanyang Avenue, Singapore 639798, Singapore}
\affiliation{Institute for Digital Molecular Analytics and Science, 59 Nanyang Drive, Singapore 636921, Singapore}

\author{Niall Byrnes\orcidlink{0000-0002-1554-3820}}
\affiliation{School of Electrical and Electronic Engineering, Nanyang Technological University, 50 Nanyang Avenue, Singapore 639798, Singapore}
\affiliation{Institute for Digital Molecular Analytics and Science, 59 Nanyang Drive, Singapore 636921, Singapore}

\author{Matthew R. Foreman\orcidlink{0000-0001-5864-9636}}
\email{matthew.foreman@ntu.edu.sg}
\affiliation{School of Electrical and Electronic Engineering, Nanyang Technological University, 50 Nanyang Avenue, Singapore 639798, Singapore}
\affiliation{Institute for Digital Molecular Analytics and Science, 59 Nanyang Drive, Singapore 636921, Singapore}

\date{\today}

\begin{abstract}
Exceptional points (EPs) have long promised enhanced sensing of physical signals, but have practically been limited by simultaneous enhancement of noise. Aligning an EP's non-analytic response with target perturbations while suppressing noise has, however, remained challenging. Here we show that fully passive, phase-tuned multi-port scattering networks enable scattering EPs with tailored anisotropic response to perturbations. We leverage projection-induced non-unitarity to realize effective non-Hermitian behavior when measuring only a subset of system ports. By formulating EP design in terms of the discriminant of the projected scattering sub-block and its directional derivatives, we give control-counting rules relating the number of programmable link phases to achievable Riemann surface topologies. We demonstrate our framework in a four-port photonic network by designing both an anisotropic EP and a Dirac-type EP with linear splitting along two parametric directions. We further suppress the global thermal drift response of a network-based sensor to a $3/2$ power-law scaling while retaining square-root sensitivity to localized signals. Since the effective non-Hermiticity arises purely from port projection, our approach transfers to integrated photonic and microwave meshes, acoustic circuits, and projected metasurfaces, offering a phase-only route to reconfigurable non-Hermitian response and noise-robust EP sensing.
\end{abstract}

\maketitle

\emph{Introduction---}Non-Hermitian systems can exhibit many unique phenomena, including unidirectional invisibility \cite{Lin2011}, the non-Hermitian skin effect \cite{Yao2018}, and coherent perfect absorption \cite{Baranov2017}, which can be exploited for advanced physical functionalities, such as directional lasing \cite{Feng2014,Hodaei2014} and topological state transfer \cite{Hong2025ChiralTransfer}. Loss or gain inherent in non-Hermitian systems moreover can foster exceptional points (EPs), which correspond to positions in parameter space where the complex eigenvalues and non-orthogonal eigenvectors of the underlying system operators become degenerate \cite{Rotter2009,ElGanainy2018,Ashida2021,Bergholtz2021}. The local topology of the associated Riemann sheets near an EP is fundamentally distinct from that of non-degenerate poles and diabolic points, affecting how the system evolves upon parametric variations \cite{Heiss2012,Dembowski2001}. Accordingly, EPs have attracted significant research effort. Spectral, or resonant, EPs (corresponding to degeneracies in the eigenfrequencies and modes of an effective Hamiltonian), were historically investigated first and have been used to, for example, control lasing thresholds and implement asymmetric optical switches \cite{Peng2014,Doppler2016}. More recently, scattering EPs, which arise as degeneracies of eigenchannels of a scattering matrix (or relevant sub-block) at a fixed real frequency, are garnering more attention as they are directly accessible through port-to-port transmission measurements in, for example, acoustic and photonic systems \cite{Ju2021, He2023ScatteringEP}. Moreover, scattering EPs can be realized in strictly lossless passive systems by performing measurements on only a subset of ports, thereby producing an effective non-unitary map as energy can leak into unobserved channels \cite{Chen2025HP}. Recent demonstrations have shown that scattering EPs can enable, for example, wavefront control and unidirectional light transport \cite{Song2021Metasurface, Feng2014b, Qin2026}.

Of the many applications of EPs investigated \cite{MiriAlu2019,Ashida2021,GWS2020, Qin2026}, EP-enhanced sensing has drawn particular attention, since the non-analytic response near a generic $p$-th order EP (EP$_p$) results in eigenvalue splitting that scales as the $p$-th root of the perturbation strength \cite{GWS2020}, suggesting an enhanced response compared to conventional linear shifts or splitting \cite{Wiersig2014}. The operational meaning of ``enhancement''  has, however, been shown to depend critically on the measurement protocol, while noise can also erase or even reverse expected gains \cite{Hodaei2017,Lau2018,Wang2020NoiseEP,Langbein2018}. This has motivated searches for architectures that suppress the response of EP based sensors to detrimental noise or signal drifts while retaining sensitivity to targeted signals. Proposals have included leveraging noise-reduction \cite{Chen2019Noise}, nonlinearities \cite{Hokmabadi2019}, stochastic-resonance \cite{Li2023PhysRevLett} and exceptional hypersurfaces in parameter space \cite{Zhang2019HypersurfaceEP}. Anisotropic (or hybrid) EP response has also been demonstrated whereby linear eigenvalue splitting is manifest for specific parametric perturbations, while the more standard non-linear scaling is retained along other generic perturbation directions in parameter space \cite{Ding2018,ZhangChan2018Hybrid,Chen2025HP}. These results suggest a potential route to realizing long promised EP sensitivity gains, but they do not yet provide a practical way to align a desired fractional power response with chosen signals or suppress sensitivity to specified noise patterns.  

In this work we demonstrate that programmable multi-port scattering networks provide a practical platform for engineering scattering EPs with tailored anisotropic response, including noise-robust sensing configurations. Working with fully passive networks at fixed frequency and using discriminant-based design constraints, we show that phase-only control of internal links enables systematic programming of anisotropic EP sensitivity within measured port subspaces, without requiring gain, loss, or material engineering. We furthermore provide explicit control-counting rules that relate the number of programmable phases to achievable response anisotropies. Although in our examples we consider a photonic network, our design framework extends naturally to other passive programmable platforms, offering a route to user-defined and reconfigurable non-Hermitian response.


\begin{figure}[t]
	\centering
	\includegraphics[width=0.9\columnwidth]{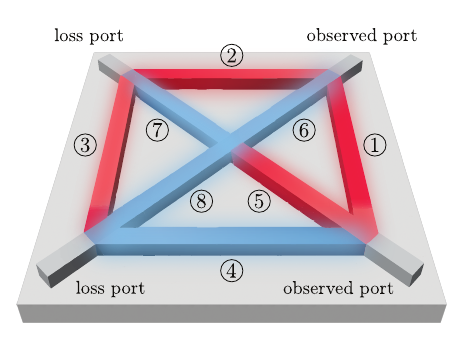}
	\caption{\textbf{Schematic of a cross-like passive programmable network.}
		Photonic waveguide network structure depicting input/output ports (gray), untuned (blue) and programmed (red) internal links. 
		\label{fig:network}}
\end{figure}


\begin{figure*}[t]
	\centering
	\includegraphics[width=\textwidth]{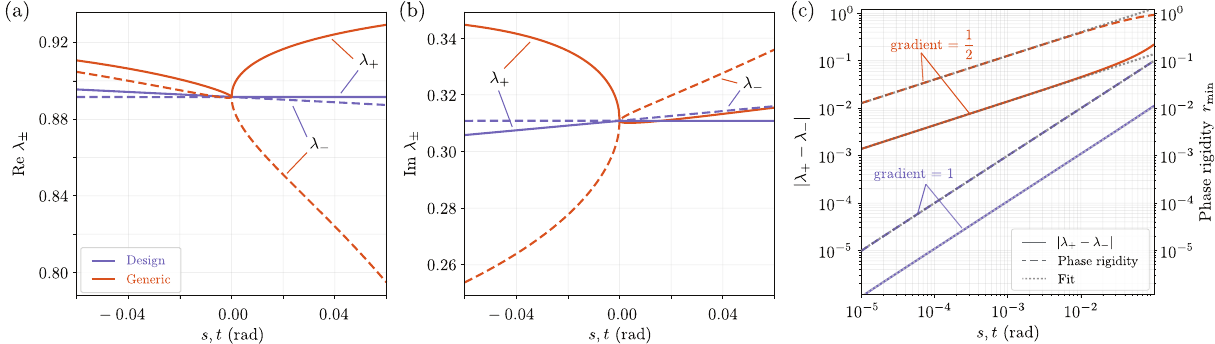}
	\caption{\textbf{Anisotropic EP in a passive programmable network.}
		(a) and (b) depict $\mathrm{Re}\,\lambda_\pm$ and $\mathrm{Im}\,\lambda_\pm$ for phase perturbations of magnitude $s$ along a generic (orange) or $t$ design (purple) direction in parameter space.
		(c) Eigenvalue separation (solid) and phase rigidity (dashed) for phase perturbations of magnitude $s$ along a generic (orange) or $t$ design (purple) direction.
	\label{fig:anisotropic}}
\end{figure*}

\emph{Anisotropic EP design framework---}We consider a fully passive multi-port wave network, or quantum graph \cite{KottosSmilansky1999,GnutzmannSmilansky2006}, with $M$ external ports and $E$ internal links (or edges), such as the one shown in Figure~\ref{fig:network}. Each link is treated as a single-mode lossless waveguide and scattering at each node is modeled assuming a Neumann boundary condition \cite{KottosSmilansky1999}. Following the graph-based scattering framework described in Ref.~\cite{ByrnesForeman2025Newton} (see also \cite{NetworkGithubRepo}), the external scattering matrix $\mathbf{S}$, which relates incoming complex field amplitudes at the $M$ external ports to the corresponding outgoing amplitudes after multiple scattering inside the graph, is obtained by matching the complex amplitudes of counter-propagating modes at each junction \cite{KottosSmilansky1999,GnutzmannSmilansky2006}. The numerical implementation is compatible with experimentally realized nanophotonic and microwave graph platforms ~\cite{Gaio2019GraphLaser, Hughes2018, Hul2004}. Throughout we work at a fixed free-space wavelength $\lambda_0$ (and wavenumber $k_0=2\pi/\lambda_0$) whereby propagation along link $\ell$ of physical length $L_\ell$ and real effective refractive index $n$ contributes a phase factor $\exp[i \phi_\ell]$, where $ \phi_\ell = n k_0 L_\ell + \theta_\ell +\epsilon_\ell$, $\theta_\ell$ is a programmable control phase shift realizable using active phase modulators \cite{Bogaerts2020, Guo2026} integrated into each link, and $\epsilon_\ell$ is a possible (e.g. environmental) perturbation or sensing signal. The vector of all control phases is denoted $\bm{\theta} = (\theta_1,\dots,\theta_E)$ with  $\bm{\epsilon}$ defined similarly for the perturbation phases. Within this model, the external scattering matrix $\mathbf{S}(k_0,\bm{\theta}, \bm{\epsilon})\in\mathbb{C}^{M\times M}$ is unitary. We partition the set of $M$ ports into measured and unobserved subsets $\mathcal{P}\subsetneq\{1,\dots,M\}$ and $\mathcal{Q} = \{1,\dots,M\} \backslash \mathcal{P}$ respectively, with the latter therefore representing effective loss channels. Consequently, the input-output mapping on $\mathcal{P}$ is described by the $m\times m$ sub-matrix $\mathbf{A}(k_0,\bm{\theta},\bm{\epsilon};\mathcal{P})$ (where $m=|\mathcal{P}|<M$) obtained by eliminating the rows and columns of $\mathbf{S}$ that correspond to ports in $\mathcal{Q}$.  Notably, $\mathbf{A}$ is generically non-unitary and can, in principle, be evaluated directly from measured scattering behavior. This projection-induced non-unitarity provides a natural, gain-free route to realizing effective non-Hermitian behavior at the level of the measured ports. 

 For this work we consider probing scattering between a set of two external ports, $\mathcal{P}_2$, whereby $\mathbf{A}(k_0,\bm{\theta},\bm{\epsilon};\mathcal{P}_2) \in \mathbb{C}^{2\times 2}$. For clarity, we henceforth drop the dependence on $k_0$ and $\mathcal{P}_2$. To approach the design and programming of anisotropic EPs we begin by considering the scalar discriminant of the characteristic polynomial of $\mathbf{A}$. Coalescence of the eigenvalues $\lambda_\pm$ of  $\mathbf{A}$ is governed by the discriminant
\begin{equation}
\Delta(\bm{\theta},\bm{\epsilon}) \equiv (\mathrm{tr}\,\mathbf{A})^2-4\det \mathbf{A}
= \bigl[\lambda_+-\lambda_-\bigr]^2 .
\label{eq:disc}
\end{equation}
 To programmatically realize an EP$_2$ in an unperturbed network structure, we can adjust the control phases $\bm{\theta}$. Specifically, we must set $\bm{\theta}=\bm{\theta}^\ast$ such that $\Delta(\bm{\theta}^\ast,\mathbf{0})=0$ and $\mathbf{A}(\bm{\theta}^\ast,\mathbf{0})$ is defective (not diagonalizable) \cite{Heiss2012}. In practice, determining a suitable $\bm{\theta}^*$ can be achieved using a numerical search algorithm. Furthermore,  since $\Delta=0$ can be achieved without defectiveness, it is necessary to additionally certify the EP numerically, which in the examples below we do by inspecting the condition number of $\mathbf{A}$ and the eigenvector degeneracy as captured by the phase rigidity \cite{Heiss2012}. Existence of a possible solution $\bm{\theta}^*$ in a general system is not guaranteed, however, for the network structures we consider we have found multiple exceptional arcs \cite{Ding2022} in $\bm{\theta}$-space for which the zero discriminant condition is satisfied.  

Realization of anisotropic response in different parametric directions requires additional considerations beyond just $\Delta=0$. To understand these extra design constraints we  consider the phase perturbation $\bm{\epsilon}$. Near a standard EP$_2$, a generic perturbation $\bm{\epsilon} = s\hat{\bm{u}}$ produces the familiar square-root splitting $|\lambda_+-\lambda_-|\propto |s|^{1/2}$, where $s\in\mathbb{R}$ is a scalar amplitude and $\hat{\bm{u}} = [u_1,u_2,\ldots,u_E]\in\mathbb{R}^E$ specifies the relative perturbation strength for each network link. We also consider a second perturbation $\bm{\epsilon} = t\hat{\bm{v}}$, ($t\in\mathbb{R}$) along a target parametric direction $\hat{\bm{v}} = [v_1,v_2,\ldots,v_E]\in\mathbb{R}^E$. This target perturbation could represent, for example, a localized  refractive index change from particle binding, formation of a fault in the network, or a known noise pattern. To program a hybrid response in which the splitting scales (say) linearly along $\hat{\bm{v}}$, while retaining the $|s|^{1/2}$ dependence for generic perturbations (i.e., anisotropic  in the sense of Refs.~\cite{Ding2018,Ding2016PRX}), we first expand the discriminant in the $\hat{\bm{v}}$ direction about the operating point viz.
\begin{align}
	\Delta(\bm{\theta}^\ast, t\hat{\bm{v}}) &\approx  t D_{\bm{v}}\Delta(\bm{\theta}^\ast,\mathbf{0}) + \frac{t^2}{2} D^2_{\bm{v}}\Delta(\bm{\theta}^\ast,\mathbf{0}) \nonumber \\
	&\quad\quad\quad\quad + \frac{t^3}{6} D^3_{\bm{v}}\Delta(\bm{\theta}^\ast,\mathbf{0})+ \ldots \label{eq:taylor}
\end{align}
where $D^n_{\bm{v}}\Delta$ is the $n$th order directional derivative of $\Delta$ along $\hat{\bm{v}}$. Since $|\lambda_+-\lambda_-| = |{\Delta}|^{1/2}$ it follows that linear eigenvalue splitting requires 
\begin{equation}
\Delta(\bm{\theta}^\ast,\mathbf{0})=0,\quad
D_{\bm{v}}\Delta(\bm{\theta}^\ast,\mathbf{0})=0
\label{eq:design_linear}
\end{equation}
and $D_{\bm{v}}^2\Delta(\bm{\theta}^\ast,\mathbf{0})\neq0$ such that the $t^2$ term is the leading non-zero term in Eq.~\eqref{eq:taylor}. 
Notably, we can add further constraints if we wish to achieve a higher order fractional power dependence, or to program the response along multiple directions in parameter space. For example, if we wish to suppress the EP response more strongly along the prescribed direction $\hat{\bm{v}}$, we can, in addition to Eq.~\eqref{eq:design_linear}, require
\begin{equation}
D^2_{\bm{v}}\Delta(\bm{\theta}^\ast,\mathbf{0})=0,\quad D_{\bm{v}}^3\Delta(\bm{\theta}^\ast,\mathbf{0})\neq0
\label{eq:design_flat}
\end{equation}
so that the leading non-zero variation of Eq.~\eqref{eq:taylor} is pushed to third order such that the eigenvalue gap scales as $t^{3/2}$. Alternatively, if we wished to obtain linear behavior along two directions $\hat{\bm{v}}$ and $\hat{\bm{w}}$, the constraints of Eq.~\eqref{eq:design_linear} could be supplemented with $D_{\bm{w}} \Delta(\bm{\theta}^\ast,\mathbf{0}) = 0$. 

Assuming that all complex constraints are independent, we are led to a simple topology independent control-counting rule. For a generic EP$_2$ in a measured $2\times2$ block, the base EP condition $\Delta=0$  first fixes a real codimension-$2$ manifold in control space \cite{Heiss2012}. Suppose now that, for all $q\in\mathbb{N}_{\geq 2}$, we demand that the eigenvalue splitting scales as $|t|^{q/2}$ for perturbations lying in the subspace spanned by the $d_q$ independent vectors  $\mathbf{v}^q_1, \mathbf{v}^q_2,\dots,\mathbf{v}^q_{d_q}$. We assume that all such perturbation directions are `generic', causing the eigenvalues to split into two distinct branches. Achieving the correct scaling for any particular vector $\mathbf{v}$ requires the vanishing of the first $q-1$ directional derivatives of $\Delta$ along $\mathbf{v}$, i.e., $D_{\bm{v}}\Delta=D_\mathbf{v}^2\Delta=\cdots=D_{\bm{v}}^{(q-1)}\Delta=0$ (and $D_{\bm{v}}^{(q)}\Delta\neq 0$), which constitutes a total of $2(q-1)$ real constraints. Realization of a suitable anisotropic EP thus requires that the number of control parameters $N$ is such that
\begin{align}\label{eq:count_general}
N \ge 2 + \sum_{q=2}^{\infty} 2 (q-1) d_q.
\end{align}
Though necessary, this condition does not guarantee the existence of an appropriate solution and different systems must be explored on a case-by-case basis. An analogous counting principle likely exists for larger measured subspaces and higher order EPs, but would require separate analysis of the relevant polynomial invariants and scaling present in the associated Puiseux expansion governing eigenvalue splitting.

\begin{figure*}[t]
	\centering
	\includegraphics[width=\textwidth]{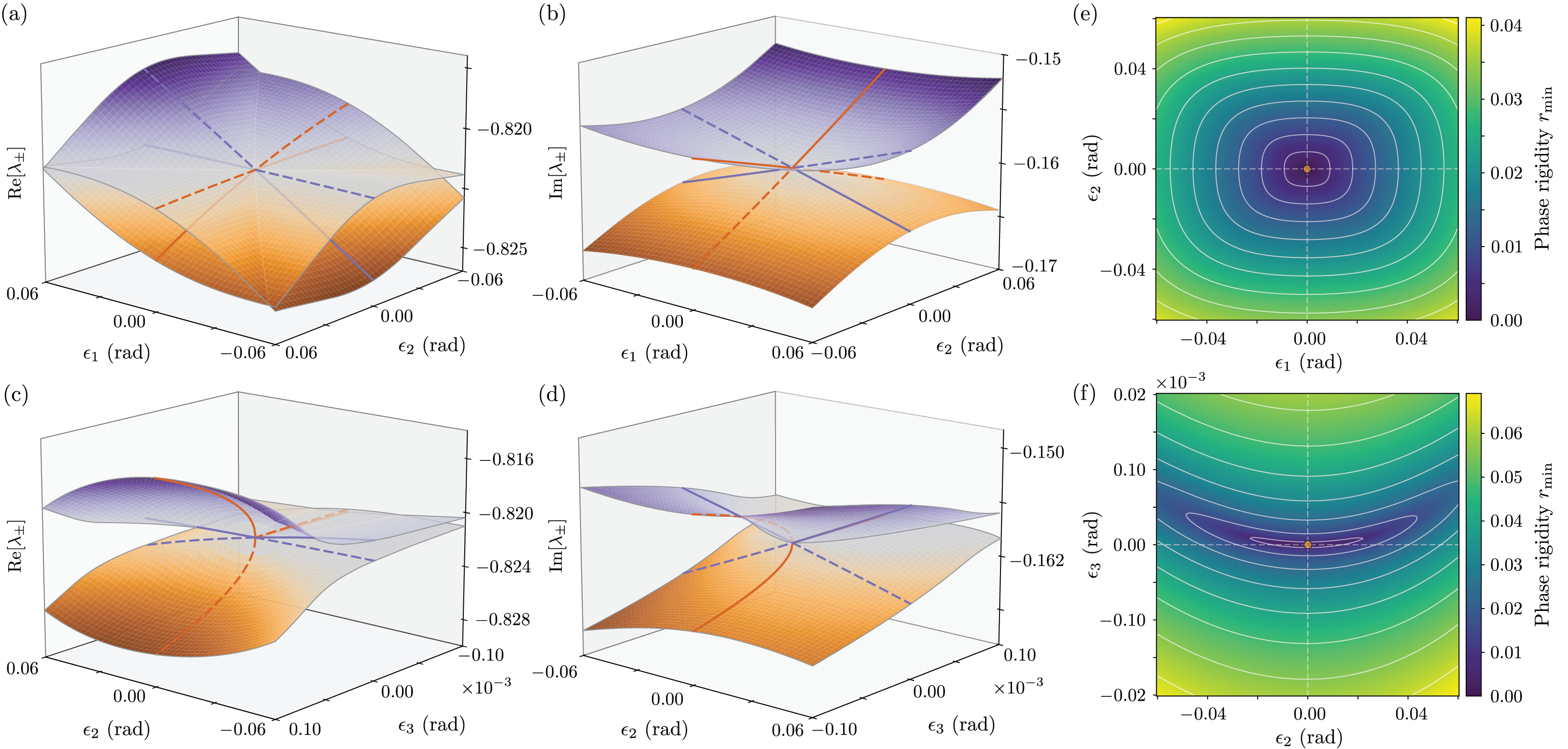}
	\caption{\textbf{Programmable hybrid response along two parametric directions.}
		(a)--(d) Real and imaginary parts of the Riemann surfaces of the projected eigenvalues $\lambda_\pm$ in the $(\epsilon_1,\epsilon_2)$ and $(\epsilon_2,\epsilon_3)$ planes. The eigenvalue gap opens almost linearly in the $(\epsilon_1,\epsilon_2)$ plane while the sheets remain connected through an exceptional point at the origin.
		(e) and (f) Phase-rigidity $r_\mathrm{min}$ over the same parametric planes, where the operating point $\bm{\theta}^\ast$ appears as a pronounced zero.}
	\label{fig:Riemann}
\end{figure*}

\emph{Example anisotropic EP designs}---We now concretely illustrate our design principles in the four-port graph shown in Fig.~\ref{fig:network}, for which $M=4$ and $E=8$, through a series of examples. The network dimension is taken such that $L_{5,6,7,8}=L_{1,2,3,4}/\sqrt{2}=100~\mu\mathrm{m}$. We further assume $n = 1.5$ and $\lambda_0 = 600$~nm.  In our first example, we demonstrate an anisotropic EP$_2$ exhibiting a linear response to phase perturbations to link~{1} ($v_j = \delta_{1j}$, where $\delta_{ij}$ is the Kronecker delta) and a square-root dependence otherwise, which we achieve through tuning the control phases of four internal links (i.e., with reference to Eq.~\eqref{eq:count_general} we have $N=4$ and $d_q = \delta_{2q}$). Specifically, we numerically search the four-dimensional phase space defined by $(\theta_1,\theta_2,\theta_3,\theta_5)$ (with all other control phases in $\bm{\theta}$ held at zero) for points satisfying the constraints of Eq.~\eqref{eq:design_linear}, and verify defectiveness for candidate points found as described above. Programmed links are indicated in red in Fig.~\ref{fig:network}. The behavior of the real and imaginary parts of the eigenvalues of $\mathbf{A}$, upon a perturbation $\bm{\epsilon}$ from the found solution $\bm{\theta}^\ast$ is shown in Figures~\ref{fig:anisotropic}(a) and (b) respectively. Purple curves show the response for phase perturbations along the design direction, i.e., to link 1 only, whereas orange curves are for a generic direction in parameter space (taken here for the sake of calculation as perturbations to link 2, i.e., $u_j = \delta_{2j}$). The corresponding eigenvalue splitting (Figure~\ref{fig:anisotropic}(c), solid lines) exhibits the expected linear and square root scaling (dotted fits). Moreover, the phase rigidity (dashed lines) shows the same scaling and critically tends to zero as $s$ and $t\rightarrow 0$ confirming eigenvector coalescence and defectiveness of $\mathbf{A}$.

Beyond exhibiting the simple anisotropic EP, our network structure, in principle, allows us to realize anisotropic responses matching different design goals. We can, as a further example, target a `higher order' anisotropic response of an EP$_2$ for which phase changes on \emph{two} specific internal links (taken here as links 1 and 2 such that $v_j = \delta_{1j}$ and $w_j = \delta_{2j}$) exhibit linear eigenvalue splitting. For our calculations a representative generic perturbation is now taken to be along $\hat{\bm{u}}$, where $u_j = \delta_{3j}$, whereby $\bm{\epsilon} = \epsilon_1\hat{\bm{v}} + \epsilon_2\hat{\bm{w}} + \epsilon_3\hat{\bm{u}}$, albeit any direction with a non-zero component orthogonal to $\bm{v}$ and $\bm{w}$ could be used. As per Eq.~\eqref{eq:count_general} we require $N=6$ control parameters  ($d_q = 2\delta_{2q}$), which in our case corresponds to independent tuning of the control phase of the links labelled 1, 2, 3, 5, 6, and 7 in Figure~\ref{fig:network}. Figures~\ref{fig:Riemann}(a) and \ref{fig:Riemann}(b) show the real and imaginary parts of the eigenvalues of $\mathbf{A}$ as we vary the phase perturbation strength in link 1 and 2. All other link phases are held constant at the found anisotropic EP point $\bm{\theta}^*$. As per the design constraints the eigenvalue gap opens approximately linearly in the $(\epsilon_1,\epsilon_2)$ plane. Notably, this Riemann topology represents a Dirac-type scattering EP \cite{Rivero2023, Wu2025}, for which the imaginary parts of the eigenvalues exhibit a conical dispersion relation (Figure~\ref{fig:Riemann}(b)) while the real parts trace hyperbolic saddle surfaces which bisect along topological branch cuts (Figure~\ref{fig:Riemann}(a)). In contrast, Figures~\ref{fig:Riemann}(c) and \ref{fig:Riemann}(d) show the Riemann surface topology for variations of $(\epsilon_2,\epsilon_3)$ from which the more familiar square-root bifurcation behavior associated with a generic EP perturbation along $\hat{\bm{u}}$ is evident. To confirm that we are indeed probing an EP of the projected scattering operator, we additionally plot the phase rigidity in Figures~\ref{fig:Riemann}(e) and \ref{fig:Riemann}(f) for the corresponding parameter spaces. A clear zero at $\bm{\theta}^*$ is observed, signalling eigenvector coalescence. Local scaling of the phase rigidity is also seen to reflect that of the eigenvalue gap, namely it varies linearly in the $(\epsilon_1,\epsilon_2)$ plane (in contrast to the square-root dependence along $\epsilon_3$).

\begin{figure}[t]
	\centering
			\includegraphics[width=\columnwidth]{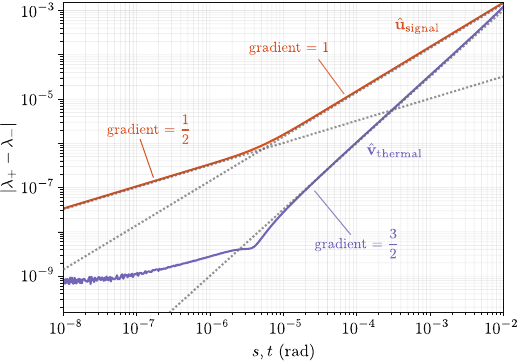}
	\caption{\textbf{Suppressed EP thermal response for sensing.} Eigenvalue splitting as a function of perturbation strength along the parametric direction $\hat{\bm{v}}_{\mathrm{thermal}}$ (purple), corresponding to a uniform temperature change across the photonic network of Figure~\ref{fig:network}, and $\hat{\bm{u}}_{\mathrm{signal}}$ (orange), corresponding to a sensing signal localized to link 3. Distinct scaling of the eigenvalue gap is evident for the different perturbations, as per the programmed EP design.}
	\label{fig:thermal}
\end{figure}
As a third and final example, we consider design of a Riemann sheet topology assuming a more physically motivated perturbation pattern relevant to a sensing context. Specifically, we consider a uniform temperature change as an undesirable environmental perturbation to the network. Temperature changes would induce a uniform refractive-index shift $\Delta n$ to each internal waveguide. Neglecting thermal expansion, for a link of length $L_\ell$ this produces a phase perturbation $\epsilon_\ell \simeq k_0\,\Delta n\,L_\ell$, implying the associated parametric drift direction is proportional to the vector of link lengths $\mathbf{L}=(L_1,L_2,\dots,L_8)$. We hence take the corresponding normalized direction, $\hat{\bm{v}}_{\mathrm{thermal}} = \mathbf{L} / \|\mathbf{L}\|$, as our design direction, for which we wish to suppress the EP response. As a representative sensing signal we choose a phase pattern  corresponding to a localized refractive index change on link~3 such that the normalized parametric shift is  $\hat{\bm{u}}_{\mathrm{signal}} = (0,0,1,0,\ldots)$.

Letting $\bm{\epsilon}=s\,\hat{\bm{u}}_{\mathrm{signal}} + t\,\hat{\bm{v}}_{\mathrm{thermal}}$, in our design we explicitly seek an $s^{1/2}$ scaling along $\hat{\bm{u}}_{\mathrm{signal}}$ and a suppressed $t^{3/2}$ scaling along $\hat{\bm{v}}_{\mathrm{thermal}}$.
Accordingly, we must impose the constraints of both Eqs.~\eqref{eq:design_linear} and Eq.~\eqref{eq:design_flat} along $\hat{\bm{v}}_{\mathrm{thermal}}$, which from Eq.~\eqref{eq:count_general} requires a total of 6 independently tunable internal links ($d_q = \delta_{3q}$). 
In other words, we require not only that $\mathbf{A}(\bm\theta^\ast,\mathbf{0})$ is an EP$_2$, but also that the leading variation of the discriminant for uniform thermal perturbations is pushed to higher order ($q=3$), while the response along the signal direction $\hat{\bm{u}}_{\mathrm{signal}}$ remains generic and therefore relatively strong (for weak signals). We show the resulting eigenvalue splitting near the found control point $\bm{\theta}^\ast$ against $s$ (orange) and $t$ (purple) in Figure~\ref{fig:thermal}. Fits to the data (using the dual-log scale shown) with their corresponding gradients are also plotted for reference. Along the design direction (purple) curves, we observe a large suppression in the system response as expected. For very small perturbations, the eigenvalue splitting observed is within the limits of our numerical precision. A residual weak $|t|^{1/2}$ scaling is also seen for perturbations in the $10^{-7}$ to $10^{-6}$ regime, which is again attributed to limited precision, in this case of our numerical search algorithm. This $|t|^{1/2}$ behavior is however rapidly superseded by the desired $|t|^{3/2}$ scaling for larger perturbations. Conversely, for the sensing signal, the usual $|s|^{1/2}$ scaling for EPs is evident at small perturbation strengths $\lesssim 10^{-5}$. Naturally, the linear scaling emerges at larger perturbations, where higher order terms in the Taylor expansion become more dominant. Notably, given the higher order suppression engineered along the noise direction, even in this regime an anisotropy of the Riemann surfaces is nevertheless maintained. From an application perspective, this example shows how programming of anisotropic EPs and the corresponding control-counting rules can translate into a concrete sensing advantage, through suppression of a drift. In a programmable multi-port platform, an effective drift vector $\hat{\mathbf{v}}_{\text{thermal}}$ can be experimentally calibrated by monitoring how observed scattering signals respond to controlled changes in temperature. The target signal pattern $\hat{\mathbf{u}}_{\text{signal}}$, meanwhile, is matched to the sensing task, such as an induced refractive index change upon analyte binding to a functionalized sensing surface. Phase-only tuning can then be employed to program an EP with suppressed response along $\hat{\mathbf{v}}_{\text{thermal}}$, while maintaining EP-enhanced sensitivity along $\hat{\mathbf{u}}_{\text{signal}}$.

\emph{Conclusion}---In this work we have shown that non-Hermitian degeneracies can be systematically programmed in a fully passive multi-port network by working in a subspace of all possible ports. Phase-only control can, specifically, be used to reshape the topology of the Riemann surfaces associated with the corresponding sub-block of the system scattering matrix, allowing realization of scattering EPs without requiring gain or loss engineering. Through formulation of the design problem in terms of the discriminant of the underlying characteristic equation, and its directional derivatives, we have established counting rules describing how the available numbers of system control parameters relate to achievable Riemann topologies. Although we have here restricted to $2\times 2$ projected blocks, such that the discriminant was a natural design metric, suitable scalar functions can be found for larger numbers of output ports. For instance, for three observed ports the scalar conditions for repeated roots of the characteristic (cubic) polynomial of $\mathbf{A}$ are well known (see e.g. \cite{Wang2025a}).  We subsequently numerically demonstrated the proposed design principle in a four-port wave network through engineering (i) a `conventional' anisotropic EP with linear eigenvalue splitting along a specified direction in parameter space, (ii) higher order anisotropy in which linear scaling is achieved along two directions, and finally, (iii) strong suppression of system response ($\sim |t|^{3/2}$) along a physically motivated noise direction. Leveraging such noise suppression could be employed to realize enhanced EP sensing. Notably, since 
in our approach, non-Hermitian behavior is achieved via projection-induced non-unitarity, the design and EP engineering framework discussed can be equally applied to microwave meshes, integrated photonic meshes, acoustic circuits, and projected metasurfaces, providing a practical route to programming non-Hermitian responses in distinct hardware platforms.

\begin{acknowledgments}
K.W. was supported by a Nanyang Technological University Interdisciplinary Graduate Program (NTU-IGP) Research
Scholarship. N.B. and  M.R.F. were supported by funding from the Institute for Digital Molecular Analytics and Science
(IDMxS) under the Singapore Ministry of Education Research Centres of Excellence scheme (EDUN C-33-18-279-V12) and by Nanyang Technological University Grant SUG:022824-00001.
\end{acknowledgments}

\end{document}